\documentclass[a4paper]{article}

\usepackage[utf8]{inputenc}
\usepackage{array}
\usepackage{bm}
\usepackage{units}
\usepackage{amsmath}
\usepackage{amssymb}
\usepackage{color}
\usepackage[english]{babel}
\usepackage{psfrag}
\usepackage{float}

\newcommand{\abs}[1]{\left\vert#1\right\vert}

\title{Constraining the fundamental interactions and couplings with
  E\"{o}tv\"{o}s experiments }
\author{ Lucila Kraiselburd$^{1}$\footnote{lkrai@fcaglp.fcaglp.unlp.edu.ar}
  and Hector Vucetich $^1$\footnote{vucetich@fcaglp.unlp.edu.ar} \\[5pt]
  $^1${\it Grupo de Gravitaci\'on, Astrof\'isica y Cosmolog\'ia,} \\ [-2pt]
  {\it Facultad de Ciencias Astron\'omicas y
    Geof\'{\i}sicas,} \\ [-2pt]
  {\it Universidad Nacional de La Plata,} \\ [-2pt]
  {\it Paseo del Bosque S/N (1900) La Plata, Argentina } \\
}

\begin{document}
\maketitle
\bibliographystyle{unsrt}

\begin{abstract}
  Upper bounds for the violation of the Weak Equivalence Principle
  (WEP) by the Fundamental Interactions have been given before. We now
  recompute the limits on the parameters measuring the strength of the
  vio\-%
  la\-tion with the whole set of high accuracy Eötvös
  experiments. Besides, limits on spatial variation of the Fundamental
  Constants $\alpha$, $\sin^2\theta_W$ and $v$, the vacuum expectation
  value of the Higgs field, are found in a model independent
  way. Limits on other parameters in the gauge sector are also found
  from the structure of the Standard Model.

  % Several authors in the past have given upper bounds for the
  % violation parameters of the Weak Equivalence Principle (WEP) due to
  % the contributions of the Fundamental Forces. We now recalculate
  % these parameters again but using the most current E\"otv\"os
  % experiments. Besides, we are putting constraints to possible spatial
  % variations of fundamental constants $\alpha$, $\lambda_{QCD}$,
  % $\sin^2\theta_W$ and $G_F$ in such way that they will be model
  % independent. Finally, we are connecting them with spatial changes of
  % gauge fundamental constants $\alpha_1$, $\alpha_2$, $\alpha_3$ and
  % the Higgs expectation value $v$.
\end{abstract}

\section{Introduction}
\label{sec:Intro}

The Standard Model of Fundamental Interactions (SM) together with
General Relativity (GR) provide a consistent description of all known
local low energy phenomena (i.e. low compared to the Grand Unified Theory
(GUT) energy scale) in good agreement with the experiment. These theories
depend on a set of pa\-ra\-%
me\-ters called the ``fundamental constants'',
which are supposed to be universal parameters; i.e.: time, position
and reference frame invariant \cite{Uzan:2002vq,Uzan:2010pm}.

The Equivalence  Principle (EP) is the physical basis of gravitational
theory \cite{Einstein1911}. 
% Briefly, it states that a freely falling  reference
% frame is equivalent to an inertial reference frame
% \cite{Einstein1911}.
There are several versions of the EP \cite{1993tegp.book.....W}. The
Weak Equivalence Principle (WEP) (also called Universality of Free
Fall (UFF)) states that the trajectory of a freely falling test body
is independent of its internal structure and composition. The Einstein
Equivalence Principle (EEP) enhances the previous version imposing the
equivalence between a local inertial reference frame and a freely falling
one.
The unrestricted validity of this very strong statement (Strong
Equivalence Principle) implies that
General Relativity is the unique theory of the gravitational field
\cite{Weinberg:Grav}. Thus, experimental tests of its validity probe
deeply into the structure of gravitation.

The traditional model for describing a WEP 
breakdown is to assume that an anomalous acceleration of body $A$,
defined as $\delta\bm{a}_A = \bm{a}_A - \bm{g}$ is due to a difference
between its inertial mass $m_A^{I}$ and its passive gravitational
mass $m_A^{P}$, i.e. the coupling constant of $A$ to the gravitational
field. It is usual to parametrize $\delta m_A$ in the form
\cite{1993tegp.book.....W}
\begin{equation}
  \label{eq:Par:PEQ:Break}
  \delta m_A = m_A^{P} - m_A^{I} = \sum_K \Gamma_K \frac{E_K}{c^2},
\end{equation}
where $E_K$ are the different contributions to the binding energy of
$A$, and $\Gamma_K$ are dimensionless parameters quantifying the
breakdown of UFF. These parameters can be estimated from Eötvös
experiments (Cf. Sect. \ref{sec:Eot:Prim}).

Most of the old published estimates have only taken into account the
binding energy contribution to the nucleus mass, which is generally
dominant.
% But nowadays, the contribution coming from the mass
% differences is also being analyzed so as to achieve a higher accuracy
However, the contribution of the binding energies of nucleons is
important for several pro\-%
blems and should be included, as it has been
done in Ref. \cite{Chamoun:1999pe}.
% (we are using the expression
% given in Ref. [6], to calculate the  parameters)} 
This generalization of the classical model (\ref{eq:Par:PEQ:Break})
will be used in a forthcoming analysis (Cf. Eq. (\ref{eq:delMass:EB})).

One of the consequences of the Equivalence Principle is that the
fundamental constants must be universal parameters, because any
dependence on time, position or reference frame would break the
equivalence with an inertial frame
\cite{1962PhRv..127..629P,Dicke:1964:ExRel}. The particular case of
space dependence of dimensionless constants has been treated in those
references and with weaker hypotheses (essentially energy
conservation)
% by Haugan
in reference \cite{1979AnPhy.118..156H}. 

In short, the binding energies  ${E_K}$ of bodies such as nuclei are
functions of the fundamental constants, and each gives a contribution
$\frac{E_K}{c^2}$ to the mass. If the fundamental constants are space
dependent, so is the mass of the body. 
In those conditions, the Lagrangian of a body in a gravitational field
%  whose mass
% is position dependent through the spatial variation of fundamental
% constants
takes the form
\begin{equation}
  \label{eq:Lagr:Eot}
  L = - \int m(\alpha) \sqrt{g_{\mu\nu}u^\mu u^\nu } ds.
\end{equation}

In the nonrelativistic limit one finds an anomalous acceleration
% \begin{equation}
%   \label{eq:Pos:Dep}
%   \delta\bm{a} = - {c^2} \frac{\bm{\nabla}m(r)}{m(r)}
% \end{equation}
% and if the position dependence is due to the space variation of the
% coupling constants $\alpha_j$ one finds
\begin{equation}
  \label{eq:Pos:Dep:alfa}
  \delta\bm{a} =  - \sum_j \frac{c^2}{m} \frac{\partial m}{\partial \alpha_j}
  \bm{\nabla}\alpha_j,   
\end{equation}
where $j$ runs over the set of fundamental constants. This anomalous
acceleration is composition-dependent and its existence can be tested
through Eötvös experiments (Sec. \ref{sec:Eot:Prim}).

Recently, the detection of a spatial variation of the fine structure
constant has been reported \cite{2011PhRvL.107s1101W,%
  2012arXiv1202.4758K,2012arXiv1203.5891B} with a gradient amplitude
$\unit[(3.6\pm0.6)\times10^{-6}]{Gpc^{-1}}$ at a $\sim 6\sigma$
level. This tantalizing result suggests that local variation of
$\alpha$ should be tested via local experiments. And the Eötvös
experiment (Section \ref{sec:Eot:Prim}) offers an excellent tool for
that. Indeed, Dent \cite{Dent:2008gu} has made such an analysis (See
also, \cite{Uzan:2010pm,2003AIPC..670..298V}).

The purpose of this paper is to analyze the space variation of the
fundamental constants in the SM, using the available Eötvös experiment
results (Cf. Table \ref{tab:Eot:Res}). We shall limit ourselves to the
Gauge sector of the Standard Model, with the exception of the vacuum
expectation value of the Higgs field. The organization is as follows:
Section \ref{sec:Eot:Prim} summarizes the main characteristics and
results of the Eötvös experiments, and describes the models we shall
use for our analysis. Section \ref{sec:BE:FC} delineates our
implementation of the structural characteristics of the test bodies
such as binding energies and constitutive relations in the Standard
Model. Section \ref{sec:Resul} shows our results and in Section
\ref{sec:Concl} we state our conclusions.

\section{A primer on Eötvös experiments}
\label{sec:Eot:Prim}

The Eötvös experiment \cite{1922AnP...373...11E}, one of the most
sensitive tests of the Equivalence Principle, measures the difference
of acceleration between two masses $A, B$ in the same gravitational
field.  It consists in suspending a pair of bodies from the arms of a
torsion balance in a homogeneous gravitational field. It is easy to
show that a differential acceleration would produce a torque
\cite{Weinberg:Grav,1993tegp.book.....W}
\begin{equation}
  \label{eq:Eot:Tor}
  T = L W \eta(A,B),
\end{equation}
where $L$ and $W$ are the lever arm of the torsion balance and the
gravitational force on the body respectively, and $\eta(A,B)$ is the
Eötvös parameter. The torsion balance is rotated with a well defined
angular velocity $\omega$ with respect to the external gravitational
field and only signals with the corresponding period are analyzed in
order to clean the result of spurious systematic effects. Additionally,
it would be possible to find any ``privileged direction'' defined by a
gradient in the masses if a nonzero result were found.

The main result of the Eötvös experiment is the Eötvös parameter
$\eta(A,B)$, defined as follows: If $\bm{g}$ is the local acceleration
of gravity,
\begin{equation}
  \label{eq:Def:EotPar}
  \eta(A,B) = \frac{(\bm{a}_A - \bm{a}_B)\bm{\cdot n}}{\abs{\bm{g}}},
\end{equation}
where $\bm{a}_{A,B}$ are the accelerations of the bodies in the
gravitational field and $\bm{n}$ a suitably chosen unit vector. Since
the Equivalence Principle implies that $\bm{a}_A = \bm{a}_B = \bm{g}$,
a non-null $\eta$ signals its breakdown. The beautiful design of the
experiment cancels many causes of error and during the twentieth century 
several orders of magnitude in accuracy have been improved. Table
\ref{tab:Eot:Res}  displays the results of several high accuracy Eötvös
experiments.

\begin{table}[tb]
  \centering
  \begin{tabular}[c]{|l|l|l|>{$}r<{$}@{$\pm$}>{$}l<{$}|c|}
    \hline
    Body & Body & Source & (\eta & \sigma)\times10^{11} & Ref. \\
    \hline
    Al & Au & Sun & 1.3 & 1.5 & \cite{RKD64}\\
    Al & Pt & Sun & 0.03& 0.04& \cite{Braginski72}\\
    Cu & W  & Sun & 0.6 & 2.0 & \cite{KeiFall82}\\
    Be & Al & Earth & -0.02 & 0.28 & \cite{1994PhRvD..50.3614S}\\
    Be & Cu & Earth & -0.19 & 0.25 & \cite{1994PhRvD..50.3614S}\\
    Si/Al & Cu & Sun & 0.51 & 0.67 &  \cite{1994PhRvD..50.3614S}\\
    Moon-Like & Earth-Like & Sun &  0.005 & 0.089 &
    \cite{1999PhRvL..83.3585B}\\ 
    Be & Ti & Earth &  0.003 & 0.018 & \cite{2008PhRvL.100d1101S}\\
    Be & Al & Earth & -0.015 & 0.015 & \cite{Adelberger:2009zz}\\
    Be & Ti & Sun & - 0.031 & 0.045 & \cite{Adelberger:2009zz}\\
    Be & Al & Sun & 0.0 & 0.042 & \cite{Adelberger:2009zz}\\
    \hline
  \end{tabular}
  \caption{Results of the Eötvös experiment}
  \label{tab:Eot:Res}
\end{table}

Equation \eqref{eq:Eot:Tor} depends crucially on the homogeneity of
the gravitational field $\bm{g}$ and great efforts have been made to
design the torsion balance so that its small
inhomogeneities are canceled. Besides, due to the design, the Eötvös
experiment is 
sensitive only to the horizontal component of the gravitational
field. Thus, only the deviation of the Earth gravitational field
$g_\perp$ or the solar gravitational field $g_\odot$ are used in the
experiments. The references cited in Table \ref{tab:Eot:Res} include
many details on the design of the experiment and the analysis of
experimental data.

Let 
\begin{equation}
  \label{eq:Mass:EB}
  M(A,Z) = m_p Z + m_n N + m_e Z - \frac{B(Z,A)}{c^2}
\end{equation}
be the atomic mass of a body of mass number $A$, atomic number $Z$,
neutron number $N = (A - Z)$ and binding energy $B(Z,A)$. The difference
between inertial and passive gravitational mass of the above body will
be \cite{Chamoun:1999pe}
\begin{equation}
  \label{eq:delMass:EB}
  \begin{split}
    \delta M &= \delta m_p Z + \delta m_n N + \delta m_e Z -
    \delta\frac{B(Z,A)}{c^2} \\
    &= \frac{\delta(m_p + m_n + m_e)}{2} A + \delta Q \frac{(N -
      Z)}{2} - \delta\frac{B(Z,A)}{c^2},
  \end{split}
\end{equation}
where
\begin{equation}
  \label{eq:Def:Q}
  Q = m_n - m_p - m_e
\end{equation}
is the decay energy of the neutron. The relative mass difference will be
\begin{equation}
  \label{eq:delMass:Frac}
  \frac{\delta M}{M} \simeq \frac{\delta(m_p + m_n + m_e)}{2m_p} +
  \frac{N - Z}{2A}\frac{\delta Q}{m_p} - \delta\frac{B(Z,A)}{Am_pc^2}.
\end{equation}
an expression which includes both the nuclear binding energy $B(Z,A)$
and the contribution of the particle rest masses $\delta m_k$. This
model is equivalent to work with constant masses for the nucleons is
some suitable system of units.

With model (\ref{eq:delMass:Frac}) (which generalizes model
\eqref{eq:Par:PEQ:Break}), the Eötvös parameter reads
\begin{equation}
  \label{eq:Eot:Expr}
  \eta(X,Y) = \frac{\delta m_X}{m_X} - \frac{\delta m_Y}{m_Y} =
  \sum_K \Gamma_K\left(\left.\frac{\hat{E}_K}{Mc^2}\right|_{X}-
  \left.\frac{\hat{E}_K}{Mc^2}\right|_{Y}\right), 
\end{equation}
where
\begin{equation}
  \label{eq:def:hatE}
  \frac{\hat{E}_K}{Mc^2} = \frac{N - Z}{2A}\frac{\delta Q_K}{m_p} -
  \delta\frac{B(Z,A)_K}{Am_pc^2}, 
\end{equation}
includes the contribution of each form of energy to the binding
energies of neutron and proton.  A set of experiments with bodies of
different compositions permits in principle the measurement of the
$\Gamma_K$ parameters.

Finally, we shall parametrize $ \frac{\hat{E}_K}{Mc^2}$
either in the generalized ``classical'' form \eqref{eq:Eot:Expr} for a test
of the Equivalence Principle, or in the form \eqref{eq:Pos:Dep:alfa}
for testing the position dependence of the fundamental constants. In
the last case, the Eötvös parameter will read, after some algebra,
\begin{equation}
  \label{eq:Eot:Par:Grad}
  \eta(A,B) = \frac{c^2}{g} \sum_j
  \frac{\partial\ln{\frac{M^{(A)}}{{M^{(B)}}}}}
  {\partial\ln\frac{\alpha_j}{\Lambda_j}}
  \bm{n\cdot\nabla}\ln\frac{\alpha_j}{\Lambda_j}
\end{equation}
where $\Lambda_j$ are suitable normalization constants.

\section{Binding energies and fundamental constants}
\label{sec:BE:FC}

The main ingredients for our analysis are the binding energies $E_K$
and their dependence on the fundamental constants. We shall discuss
separately nuclear binding energies and neutron-proton mass differences.

\subsection{Nuclear binding energies}
\label{ssec:NucBE}

The largest contribution to the binding energy of an atom comes from
the nuclear binding, which has been discussed for a long time.  The
simplest approach is to use the semi-empirical mass formula
\cite{Weiszaecker35,martin09} complemented with the estimate of the
weak interactions contribution to the binding energy
\cite{HauWilWI,FishbachWI}. There are simple analytic approximations
for the strong, Coulomb and weak contributions to the binding energy
$B$, namely
\begin{subequations}\label{sub:BEs}
  \begin{gather}
   \frac{E_S}{Mc^2} = a_V -  a_S A^{-\nicefrac{1}{3}} -
   a_A\frac{(N-Z)^2}{A^2}, \label{eq:BES}\\
  \frac{E_C}{Mc^2} = \frac{3e^2}{5r_0}
  \frac{Z(Z-1)}{A^{\nicefrac{4}{3}}}, \label{eq:BEC} \\
  \begin{split}
     \frac{E_W}{Mc^2} =&  \mathcal{G} G_F2^{-2/3}V^{-1}
     \left\{ NZ \left[(3\alpha_{\beta}^2-1)- 4\left(\frac{1}{2} -
           2*\sin^2\theta_W\right) \right] \right.\\ 
           &+ \left. \frac{N^2}{2} +
           \frac{Z^2}{2} (4\sin^4\theta_W - 2\sin^2\theta_W + 1)
         \right\}.  
    \end{split} 
  \end{gather} \label{eq:BEW}
\end{subequations}

In the above equations $r_0A^{\nicefrac{1}{3}}$ is the nuclear radius,
$V = \nicefrac{4\pi}{3} A r_0^3$ the nuclear volume and $N = A -Z$ the
neutron number. $\alpha_{\beta}$ is the $\nicefrac{G_A}{G_V}$ ratio
for neutron decay, and $G_F$ and $\theta_W$ are the Fermi constant and
the Weinberg angle respectively. $\mathcal{G}$ is an enhancement
factor of the weak interactions due to the strong ones
\cite{FishbachWI}. 
Besides, the ``strong constants''
$a_V, a_S, a_A$ as well as $m_p, m_n$ are all proportional to
$\Lambda_{\rm QCD}$ in the chiral limit.

These  analytic expressions, which are reasonably accurate,
display the dependence of the nuclear binding energies on the
fundamental constants.

\subsection{Neutron-proton mass difference}
\label{ssec:NP:MassDif}

The other contribution to the mass is the neutron-proton mass
difference which contributes to the neutron decay energy $Q$. 

Model independent contribution of the strong, electromagnetic and weak
forces to the neutron-proton mass difference $\Delta M$ can be
computed with the Cottingham formula \cite{Cottingham:1963zz} and its
generalizations for the strong \cite{Epele91} and weak
\cite{Chamoun:1999pe} interactions. Their calculated values are:
\begin{align}\label{ecs:NP:SEW}
  \left.\frac{\Delta M}{M}\right|_S &= 2.22\times10^{-3}, &
  \left.\frac{\Delta M}{M}\right|_E &= -0.83\times10^{-3}, &
  \left.\frac{\Delta M}{M}\right|_W &= -5.0\times10^{-9}.
\end{align}

However, the explicit dependence on the fundamental constants is not
obvious. A careful analysis of the respective expressions shows that
the electromagnetic contribution is proportional to $\alpha$ and the
weak one to $G_F$. Besides, the weak contribution has a dependence on
$\sin^2\theta_W$, which must be numerically computed with reference
\cite{Chamoun:1999pe} method. The result is
\begin{equation}\label{eq:NP:s2w}
  \frac{\sin^2\theta_W}{M}\frac{\partial \Delta
    M}{\partial\sin^2\theta_W} 
  \simeq  2.0\times10^{-8}.  
\end{equation}

Finally, an important result is that the ``strong'' contribution to
$\Delta M$ is not proportional to $\Lambda_{\rm QCD}$ near the chiral
limit but to the $u-d$ quarks mass di\-ffe\-%
rence, a result that can be
derived in an elementary way from Chiral Perturbation Theory
\cite{1982PhR....87...77G} and that is quantitatively confirmed in
lattice calculations (See, for instance, \cite{Beane:2006fk}).

Since quark and electron masses are proportional to  the vacuum
expectation value of the Higgs field  $v$, $m_i = y_i v$, so is
$Q$. The available Eötvös experiments are not enough to separate the
Yukawa coupling parameters and $v$. So in this paper we limit ourselves
to the analysis of the gauge sector plus the single Higgs sector
parameter $v$. With this limitation, we find the following 
expression for $Q$ as a function of the fundamental constants
$\alpha$, $v$ and $\sin^2\theta_W$:
\begin{equation}
  \label{eq:Q:FC}
  \frac{\delta Q}{M} = \frac{\delta\alpha}{\alpha}\left.\frac{\Delta
      M}{M}\right|_E + \frac{\delta v}{v}\frac{Q}{M} + \frac{\sin^2\theta_W}{M}\frac{\partial
    \Delta M}{\partial\sin^2\theta_W}
  \frac{\delta\sin^2\theta_W}{\sin^2\theta_W} 
\end{equation}

If for each of the fundamental constants $\alpha_i$ we replace
\begin{equation}
  \label{eq:Spa:Var}
  \delta\alpha_i = \bm{\nabla\alpha}_i\bm{\cdot}\delta\bm{r}
\end{equation}
we obtain the contribution of $Q$ to the Eötvös parameter.

\subsection{Constitutive relations}
\label{ssec:St:Mod}

In this subsection we shall use the fine structure constant $\alpha$,
the vacuum expectation value of the Higgs field $v$ and the squared sine
of Weinberg's angle $\theta_W$ as our basic variables. Other
fundamental constants from the gauge sector are related to our basic
constants in the form
\begin{subequations}\label{ecs:Cte:Gauge}
\begin{align}
  \alpha &= \alpha_1 \sin^2\theta_W, & \tan^2\theta_W &=
  \frac{\alpha_2}{\alpha_1}, &
  G_F &= \frac{1}{\sqrt{8}v^2}, \\ 
M^2_W &= \frac{\alpha_1}{2}v^2, &
  M^2_Z &= \frac{ M^2_W}{\cos^2\theta_W}, & \alpha_3 &=
  \frac{\beta^{-1}}{\ln\frac{\mu^2}{\Lambda^2_{\rm QCD}}}. 
\end{align}
\end{subequations}

The last equation shows that in  QCD system of units, $\alpha_3$ is
automatically constant.

\subsection{Scaling and systems of units}
\label{ssec:Scal:Units}

The need of working with nondimensional quantities when studying the
variation of fundamental constants it has been discussed in many
papers (See, for instance, \cite{Uzan:2002vq,Uzan:2010pm}) since a
suitable choice of units may cancel its variation. Many measurements,
however, are carried out on dimensional quantities and its a\-na\-%
ly\-sis must be done starting with these data. This problem may be
solved either by transforming the dimensional quantities to a standard
system of units \cite{1990PhRvD..41.1034S} or transforming these
dimensional quantities into dimensionless ones through division by a
suitably chosen constant.

One of the beauties of the Eötvös experiment is that it has a
``natural'' way of defining $\eta$ as
nondimensional parameter, and equation \eqref{eq:Eot:Expr} is already in
dimensionless form. Besides, since the anomalous acceleration
\eqref{eq:Pos:Dep:alfa} can be written as
\begin{equation}
  \label{eq:Pos:Dep:alfa:2}
  \delta\bm{a} =  - \sum_j \frac{c^2}{m} \frac{\partial m}{\partial \alpha_j}
  \bm{\nabla}\ln\alpha_j. 
\end{equation}
Any  normalization constant will not contribute to
the differential acceleration. 

In this paper we shall use ``QCD units'': that is, we assume that
\begin{equation}
  \label{eq:Def:QCD:Unit}
  \bm{\nabla}\Lambda_{\rm QCD} = 0.
\end{equation}

In the nuclear binding energies the dependence of the strong binding
energy on the $(u, d, s)$ quark masses can be neglected. This implies
that $\Gamma_S = 0$ since we are working near the chiral limit.

However, above approach is not always correct. Indeed, the logarithmic
derivatives of the binding energy parameters $a_V, a_S, a_A$ with
respect to the quark masses could be large in some cases.
Refs. \cite{2010PhRvD..82h4033D,2010CQGra..27t2001D,2011CQGra..28p2001D}
make a detailed analysis on the subject. In our case, the only
important contribution from the quark masses is in the ``strong''
contribution to the $p-n$ mass difference.

\section{Results}
\label{sec:Resul} 

We have performed the above sketched calculations both to test the
Equivalence Principle and the existence of gradients of the
fundamental constants in the Standard Model. A weighted least squares
procedure was applied to the values of $\eta$ in Table
\ref{tab:Eot:Res},  the conditional equation being given by either
equation \eqref{eq:Eot:Expr} or in the form corresponding to spatial
variation \eqref{eq:Eot:Par:Grad}. 

\subsection{Test of the Equivalence Principle}
\label{ssec:Test:EP}

\begin{table}[tb]
  \centering
  \begin{tabular}[c]{|>{$}l<{$}|>{$}r<{$}@{$\pm$}>{$}l<{$}||rrr|>{$}c<{$}|}
    \hline
    \text{Param} & \Gamma & \sigma(\Gamma)
    &\multicolumn{3}{c|}{Correlations} & \text{Up. Bound}\\ 
    \hline
    \Gamma_S & (-1.3 & 1.6)\times10^{-10} & 1.00 & 0.98 & 0.84 &
    5\times10^{-10}\\ 
    \Gamma_C & (-3.0 & 4.0)\times10^{-10} & 0.98 & 1.00 & 0.87 &
    1.2\times10^{-9}\\ 
    \Gamma_W & (-2.7 & 8.6)\times10^{-4}  & 0.85 & 0.87 & 1.00 &
    2.6\times10^{-3}\\ 
    \hline
    \Gamma_C & (1.5 & 6.7)\times10^{-11}  &      & 1.00 & 0.00 &
    2\times10^{-10}\\ 
    \Gamma_W & (2.2 & 2.0)\times10^{-4}   &      & 0.00 & 1.00 &
    6\times10^{-4}\\ 
    \hline
  \end{tabular}
  \caption{Test of the Equivalence Principle.  }
  \label{tab:Eq.Pr:Test}
\end{table}

Table \ref{tab:Eq.Pr:Test} shows our results for the test of the
Equivalence Principle. We have used the expressions \eqref{sub:BEs} with
the contributions \eqref{ecs:NP:SEW} from the neutron-proton mass
difference. The first three lines of the table show the result
assuming that the three interactions break the Equivalence
Principle. The last two lines assume that the strong contribution
satisfies the Equivalence Principle ($\Gamma_S = 0$). An enhancement
factor $\mathcal{G}  = 8$ for the Weak Interactions has been assumed
\cite{Chamoun:1999pe}. Our results are similar to those of reference
\cite{Chamoun:1999pe}, but the new bounds are smaller due to the
inclusion of higher accuracy results
\cite{1999PhRvL..83.3585B,2008PhRvL.100d1101S,Adelberger:2009zz}. 

The large correlations in the first three lines suggest that either
the breakdown of the Equivalence Principle should be analyzed
simultaneously or that a constraint such as ($\Gamma_S = 0$) should be
imposed and in this case only violation parameters relative to the
strong interactions will be found.

\subsection{Spatial variation of Fundamental constants}
\label{ssec:Spa:Var:SM}

\begin{table}[tb]
  \centering
  \begin{tabular}[c]{|>{$}l<{$}|>{$}r<{$}@{$\pm$}>{$}l<{$}||rrr|>{$}c<{$}|}
    \hline
    \text{Param} & \Gamma & \sigma(\Gamma) &\multicolumn{3}{c|}{Correlations} &
    \text{Up. Bound}\\
    \hline
    \Theta_\alpha & (0.5 & 1.6)\times10^{-10} & 1.00 & 0.74 & 0.71 &
    4.8\times10^{-10}\\
    \Theta_v & (-0.5 & 2.5)\times10^{-8} & 0.74 & 1.00 & 0.99 &
    1.1\times10^{-7}\\
    \Theta_{\sin^2\theta_W} & (0.1 & 1.1)\times10^{-2} & 0.71 & 0.99 &
    1.00 & 3.3\times10^{-2}\\
    \hline
  \end{tabular}
  \caption{Bounds on the space variation of fundamental constants:
    Basic va\-ria\-%
    bles.}
  \label{tab:spa:var:Bnd}  
\end{table}

\begin{table}[tb]
  \centering
  \begin{tabular}[c]{|>{$}l<{$}|>{$}c<{$}|>{$}c<{$}||>{$}l<{$}|>{$}c<{$}|>{$}c<{$}|}
    \hline
    j & \Theta_j & \ell_j \unit{pc} & j & \Theta_j & \ell_j \unit{pc} \\
    \hline
    \alpha & 4.8\times10^{-10} & 1.1\times10^{12} & v &
    7.5\times10^{-8} & 6.6\times10^9 \\
    \sin^2\theta_W & 3.4\times10^{-2} & 1.4\times10^4 & \alpha_1 &
    3.4\times10^{-2} & 1.4\times10^4\\
    \alpha_2 & 1.0\times10^{-2} & 4.8\times10^4 & G_F &
    1.5\times10^{-7} & 3.4\times10^9\\
    M_W^2 & 3.4\times10^{-2} & 1.4\times10^4 & M_Z^2 & 2.4\times10^{-2} &
    2.0\times10^4\\
    \hline
  \end{tabular}
  \caption{Bounds on the space variation of fundamental constants: Gauge sector.}
  \label{tab:spa:var:Gauge}
\end{table}

Turning to the spatial variation problem, it is convenient to work
with the nondimensional quantity \cite{2003AIPC..670..298V}
\begin{equation}
  \label{eq:Def:Theta}
  \Theta_j = \frac{c^2}{g_\odot} \frac{|\bm{\nabla}\alpha_j|}{\alpha_j},
\end{equation}
which is the ``natural'' nondimensional parameter for this problem. As
explained before, we use as basic variables $\alpha$, $v$ and
$\sin^2\theta_W$. Again, the values of the $\Theta$ parameters were
found by least squares adjustment and upper bounds were obtained as
$3\sigma$ values. The results of the adjustment are displayed in Table
\ref{tab:spa:var:Bnd} in the same format as the one in Table
\ref{tab:Eq.Pr:Test}. 

Logarithmic differentiation of the Standard Model relations in
\eqref{ecs:Cte:Gauge}, after normalization by division by suitable
powers of $\Lambda_{\rm QCD}$, yields a system of linear equations for the
gradients of the parameters from which the upper bounds of Table
\ref{tab:spa:var:Gauge} are found. The quantities
\begin{equation}
  \label{eq:Def:ell}
  \ell_j = \frac{\alpha_j}{|\bm{\nabla}\alpha_j|}
\end{equation}
define distance scales where the spatial variation of a given
fundamental constant becomes important. 

The results summarized in Tables \ref{tab:Eq.Pr:Test} to
\ref{tab:spa:var:Gauge} are the main results of this paper.

\section{Conclusion}
\label{sec:Concl}

The results stated in Sections \ref{sec:Eot:Prim} and \ref{sec:Resul}
show that no violation of the Equivalence Principle is observable in
laboratory experiments  down to the $10^{-13}$ level. The classical
model decomposition of the Eötvös parameter \eqref{eq:Eot:Expr} shows
that the contributions of the fundamental interactions to such a
violation are extremely small. 

On the other hand, the order of magnitude of the upper bounds for the
gradients of the fundamental constants are very variable, from
$\sim\unit[10^{-12}]{pc^{-1}}$ for $\alpha$ to
$\sim\unit[10^{-4}]{pc^{-1}}$ for $\alpha_2$. These extremely small
gradients of galactic or cosmological scale, are the best available
bounds on the spatial variation of fundamental constants.

The results of this paper are in a certain sense complementary to
those of reference \cite{Dent:2008gu} where the analysis was focused
mainly on the Higgs sector of the Standard Model and on the
sensitivity to the Newtonian potential. See also reference
\cite{Uzan:2010pm} for a more complete analysis of that sector.

Our upper bounds, however, are too big for an independent test of the
reported cosmological gradient of $\alpha$. Our smallest bound is
obtained assuming that only $\alpha$ has a sensible variation
\begin{equation}
  \label{eq:grad:alpha:only}
  \frac{|\bm{\nabla}\alpha_j|}{\alpha_j} <
  \unit[2\times10^{-4}]{Gpc^{-1}}, 
\end{equation}
and it is about 60 times greater than the detected one. This is not
far from the needed sensitivity and the proposed \texttt{MICROSCOPE}
\cite{Microscope} or \texttt{STEP} \cite{STEP} experiments,
whose accuracy is about a thousand times greater  should be able to
detect it.

\section*{Acknowledgments}

The authors are indebted to Drs. P. D. Sisterna and S. J. Landau for
their helpful comments and advice, and to an anonymous correspondent
for his criticisms and suggestions.

\bibliography{CoupAdj}
\end{document}